\documentclass[a4paper,12pt]{article}
\usepackage{amssymb}
\usepackage{graphicx}
\usepackage{epsfig}
\newcommand{\beq}{\begin{equation}}
\newcommand{\eeq}{\end{equation}}
\newcommand{\beqa}{\begin{eqnarray}}
\newcommand{\eqa}{\end{eqnarray}}

\title{\textbf{A SIMPLE CHAOTIC NEURON MODEL:
STOCHASTIC BEHAVIOR OF NEURAL NETWORKS}}

\author{\centerline{Ekrem Ayd\i ner\thanks{Address correspondence to Dr. Ekrem Aydiner, Department
of Physics, Faculty of Arts and Sciences, University of Cukurova,
Adana, Turkey. E-mail: ekol@cu.edu.tr}, Adil M. Vural$^1$,Bekir
\"{O}z\c{c}elik$^1$, Kerim K\i yma\c{c}$^1$} \\
\centerline{and \"{U}ner Tan $^2$}\\ $^1$ Department of Physics,
Faculty of Arts \\and Sciences, University of Cukurova, Adana,
Turkey
\\ $^2$ Department of Physiology, Medical School,\\
Cukurova University, Adana, Turkey}

\begin{document}

\maketitle

\section*{Abstract}

We have shortly reviewed the occurrence of the post-synaptic
potentials between neurons, the relation between EEG and neuron
dynamics, as well as methods of signal analysis. We supposed a
simple stochastic model representing electrical activity of
neuronal systems. The model is constructed using the Monte Carlo
simulation technique. The results yielded EEG-like signals with
their phase portraits in three-dimensional space. The Lyapunov
exponent was positive, indicating a chaotic behavior. The
correlation dimension of the EEG-like signals was found to be
0.92, which was smaller than those reported by others.  It was
concluded that this neuron model may provide valuable clues about
the dynamic behavior of neural systems.

Keywords: neurons, neuron dynamics, EEG, fractals, chaos.

Page Heading : neuron model

\section{Introduction}

It is well known that there are about one hundred billions of
neurons in the human brain each interacting with about tens of
thousands of other neurons via synapses. The complex processes
such as learning, remembering, and thinking are based on neurons
exhibiting complex interactions. Although the neurons in brain
seem to be relatively simple structures, their dynamics with
consequent integrative functions could not be understood clearly.

The studies in neurosciences are concentrated in two categories.
Studies in the first category dealt with modelling of
single-neuron dynamics, which are captured with biologically
inspired neuron model, such as those from \cite{17} and
Fitzhugh-Nagumo equations (\cite{9}, \cite{25}). However, formal
neurons, used in artificial neural networks like the \cite{24} or
the graded response neurons \cite{14} have only trivial, i.e.,
convergent dynamics as single elements. Researchers in the second
category are interested in neuronal ensembles. Their purpose is to
study the modelling of brain-like functions, and to account for
collective or clustering actions in brain
\cite{28,18,11,13,21,19}. However, chaotic neuron models based on
biological motivation were also studied \cite{2,3}.

Information processing in brain results from the spread and
interaction of electrical and chemical signals within and among
neurons. This involves nonlinear mechanisms that span a wide range
of spatial and temporal scales and are constrained to operate
within the intricate anatomy of neurons and their interconnections
\cite{8}. The mathematical equations describing the brain
mechanisms generally do not have analytical solutions, and are not
a reliable guide to understand the mechanism of the neuron
dynamics underlying the brain functions. In order to explain
neuron dynamics, models constructed to explain the brain functions
have to exhibit the nonlinear properties of the brain \cite{22}.
For this reason, the generally used linear approaches are severely
limited. Therefore, using simulation techniques would be important
in this field \cite{16}.

On the other hand, it is generally accepted that there is a
relationship between EEG signals and neuronal activity.  Analysis
of the EEG records enables us to obtain some information about the
existence of chaotic attractors, different responses to different
stimulations, epileptic seizures etc \cite{27,23,1,20,5,10}. In
the following Section, we reviewed the cycle of neuronal
post-synaptic potentials and relations of neuron dynamics to EEG
signals; we have suggested a simple neuron model exhibiting the
collective behaviors organized similar to neurons in human brain.
Although our model has no direct biological and chemical basis, we
believe that its importance lies in the fact that it is based on
the nature of stochastic electrical activity between neurons.
Finally in the last section, we analyzed collective dynamics of
neurons by using Monte Carlo method.

\section{BASIS OF NEURON DYNAMICS}

\subsection{Cyclic Post Synaptic Potentials (PSP)}
Below, we will summarize how a single neuron's PSP cycle occurs in
five steps \cite{4}. First, a neuronal axon is bistable; it can be
either active or passive. In the first state, according to the
result of the summation performed in the soma it propagates a
signal - a spike, or an action potential (AP). The shape and the
amplitude of the propagated signals are stable and replicated at
the branching points of the axon. The amplitude of this propagated
signal is of the order of tens of millivolts. In the second state,
there is no signal travelling in the axon, rather there is a
resting potential. Secondly, when the travelling signal arrives at
the end of the axon, it causes a release of neuro-transmitters
into the synaptic cleft. Thirdly, neuro-transmitters arrive at the
membrane of the post-synaptic membrane. On the post-synaptic side,
these neuro-transmitters bind to the receptors causing the latter
to open up and allow ionic currents to penetrate into the
post-synaptic neuron. The amount of penetration of ionic currents
per pre-synaptic spike is a parameter, which specifies the
efficacy of the synapse. Fourthly, PSP, contrary to the action
potential, propagates in a declining manner towards the soma,
where the inputs from all of the pre-synaptic neurons connected to
the post-synaptic neuron are integrated. Fifthly, if the sum of
the PSPs arriving within a short period exceeds a certain
threshold, the level at which the post-synaptic membrane becomes
unstable against depolarizing ionic current flows, the probability
for the emission of a spike becomes significant. This threshold is
approximately ten millivolts. A large number of inputs are
required to produce an action potential (spike).

The cyclic-time of a biological neuron in the cerebral cortex,
i.e., the time interval from the start of a spike in the
pre-synaptic neuron to the occurrence of that spike in the
post-synaptic neuron is about 1-2 milliseconds. This is the
duration for travelling of the spike to the pre-synaptic axon till
the neurotransmitter crosses the synaptic gap. Following the
dramatic event of the emission of a spike, the neuron needs time
to recover. There is a period of 1-2 milliseconds in which the
neuron cannot fire a second action potential, no matter how large
the depolarizing potential might be. This period is called the
absolute refractory period, determining the maximal discharge rate
of a neuron at about 500-1000 Hz. Such high frequencies can occur
in sensory neurons, since the stimulus is externally determined
and can be arbitrarily strong. In neurons of the cerebral cortex,
the neuronal firing rates are rather low, such as 150 Hz; it may
even be as low as 30-40 Hz \cite{4}.

\subsection{Relation Between Neuron Dynamics and EEG Signals}

During physiological processes, thousands of neurons communicating
with each other produce an overall electrical activity in the
brain, based upon the postsynaptic neuronal activities. However,
it is impossible to observe the electrical activities of a single
neuron in EEG, reflecting the overall activity of the brain. That
is, EEG signals do not give any information directly relating to
activity of single cerebral neurons. It is also assumed that the
recorded EEG signals represent the postsynaptic neurological
activities as a function of time, corresponding to a time series.
Hence, the analysis of time series of EEG signals plays a crucial
role in understanding the collective dynamics of neurons.

\subsection{Analysis of EEG signals}

It can be assumed that the system of interest, i.e., an EEG
signal, can be described by N variables, where N is a large
number, so that at any instant of time there is a point $ X\left(
t \right) = \left\{ {X_1 (t),X_2 (t),.......,X_N (t)} \right\}$ in
an N-dimensional phase space that completely characterizes the
system. Trajectory of these points in the phase space make-up an
attractor. For chaotic motion this attractor is often an object
that can be described by a fractal dimension. Such attractors are
called strange attractors.

Time series captured from the EEG signals may seem to have
one-dimensional information but in fact they may have more
dimensional information. Reconstructing this kind of series in the
phase space will show the details of the one-dimensional
information of these series. Therefore, the proposed method is the
construction of a strange attractor, and the evaluation of the
correlation dimension of this attractor.

 For sufficiently long times $\tau$, one uses the embedding theorem of Takens \cite{29}
 to construct the sequence of the displaced time series $\left\{ {\zeta \left( t \right),\zeta \left( {t + \tau }
\right),...,\zeta \left( {t + \left( {m - 1} \right)\tau }
\right)} \right\}$. This set of variables has been shown to
possess the same amount of information as the N-dimensional phase
space, provided that $ m \ge 2d + 1$,  where d is the dimension of
the attractor for the original N-variable system. The condition on
the embedding dimension m is often overly restrictive and the
reconstructed attractor usually does not require m to be so large
\cite{30}.

For constructing the phase portrait of the EGG signals on the
phase space, generally choosing m=3 is enough. Hence, one can
obtain three-dimensional construction of EEG signals as $\left\{
{\zeta \left( t \right),\zeta \left( {t + \tau } \right),\zeta
\left( {t + 2\tau } \right)} \right\}$.  It is expected that this
method produces a strange attractor corresponding to the EEG
signals, and the correlation dimension of this attractor can be
evaluated.

A more efficient approach to calculate the correlation dimension,
$D_2$, is introduced by Grassberger and Procaccia \cite{12}. This
approach is based on the behavior of a so-called correlation sum
(or correlation integral).$D_2$ has been widely used to
characterize chaotic attractors.

To define the correlation dimension, we first let a trajectory (on
an attractor) evolve for a long time, and collect as data the
values of N trajectory points. Then for each point i on the
trajectory, we ask for the relative number of trajectory points
lying within the distance r of the point i, excluding the point i
itself. Call this number $N_i (r)$. Next, we define $p_i (r)$ to
be the relative number of points within the distance r of the i th
point: $p_i \left( r \right) = N_i /\left( {N - 1} \right)$.
Finally, we compute the correlation sum $C(r)$ \cite{15,6,26}:

\begin{equation}\label{1}
C\left( r \right) = \frac{1}{N}\sum\limits_{i = 1}^N {p_i \left( r
\right)}
\end{equation}
where,
\begin{equation}\label{2}
p_i \left( r \right) = \frac{1}{{N - 1}}\sum\limits_{j = 1,j \ne
i}^N {\Theta \left( {r - \left| {x_i  - x_j } \right|} \right)}
\end{equation}

$p_i$ itself can be written in more formal terms by introducing
the Heaviside step function $\Theta$ \cite{7}:

$$
\Theta \left( x \right) = \left\{ {\begin{array}{*{20}c}
   {0x < 0}  \\
   {\begin{array}{*{20}c}
   {{\raise0.5ex\hbox{$\scriptstyle 1$}
\kern-0.1em/\kern-0.15em
\lower0.25ex\hbox{$\scriptstyle 2$}}} & {x = 0}  \\
\end{array}}  \\
   {1x > 0}  \\
\end{array}} \right.
$$

Note that $C(r)$  is defined such that $C(r)=1$  if all the data
points fall within the distance r of each other. If r is smaller
than the smallest distance between trajectory points, then $p_i=0$
for all i , and $C(r)=0$.

The correlation dimension $D_2$ is defined to be the number that
satisfies

\begin{equation}\label{3}
C\left( r \right)\sim \mathop {\lim }\limits_{r \to 0} r^{D_2 }
\end{equation}

After taking logarithms (\ref{3}) gives

\begin{equation}\label{4}
D_2  = \mathop {\lim }\limits_{r \to 0} \frac{{\log C(r)}}{{\log
r}}
\end{equation}

If two nearby trajectory on a chaotic attractor start off with a
separation of $d_0$ at a time $t=0$, then the trajectories diverge
so that their separation at time $t$ can be given as, $d\left( t
\right) = d_0 \exp \left( {\lambda t} \right)$ or for a discrete
time scale as,

\begin{equation}\label{5}
d_n  = d_0 \exp \left( {\lambda n} \right)
\end{equation}
where $\lambda$ is a measure of the divergence of nearby
trajectories. If $\lambda$ is positive, two trajectories diverge
and thus, the behavior of the trajectories is chaotic.

\section{DEFINITION OF OUR MODEL }

In this study, a simple stochastic neuron model has been
developed. The model has neither a biological nor a chemical
motivation at all. Considering this statement, we can  say that
the model does not encounter a biological neuron system such as
that in a brain. Our model can be briefly defined as follows: i)
The model consists of N pieces of neuron-like (NL) elements; ii)
Each NL element is randomly connected with other NL elements, and
the connection numbers are not equal and are assumed to be in the
interval $\sqrt N $ to $N/2$ . In other words, the connection
numbers are generated randomly in this interval; iii) a threshold
potential is defined for each NL element, and when the amount of
the accumulated signal on a NL element is above the threshold for
a single spike. After propagating the spike, the NL element
integrates the signals from the other NL elements until they reach
the lateral spike propagation criteria for a second spike and so
on. The time interval between the two propagations is random; iv)
spike propagation from the NL element complies with the PSP cycle
in principle as mentioned above. However, since our model has no
biological and chemical motivation, the intracellular chemical
cyclic mechanisms necessary for the occurrence of PSP; v) we will
define PSP cycle of each NL element as PSP-like cycle; vi) we can
state that the PSP cycle in biological neurons does not only
depend on the intracellular processes but also on the quality and
quantity of the inputs integrated from other neurons. Therefore,
it is not possible to predict the propagation of a spike from the
PSP cycle of a neuron. In other words, it is assumed that each
neuron realizes its PSP cycle stochastically. Hence, the
information necessary to understand the dynamics of an ensemble of
neurons is the time evolution of the propagation of spikes in the
system rather than the mechanisms underlying the neurons dynamics.
With the consideration of this idea, in our NL system, to
understand the dynamics we have just focused on the stochastic
behaviors of the NL elements; vii) we have analyzed the dynamics
of neurons within discrete time scales. In advance, at each
discrete value of t, only the NL elements that comply with the
PSP-like cycle participates to the propagation of spikes; viii) at
each discrete value of t, the number of NL elements propagating
spikes is assumed to be the scale of the integrated signals
quantity that this NL system generates. Therefore, we assumed -as
the amount of the integrated signals- that belong to the whole
system to be as the amount of integrated signals that were
generated by the NL elements within the system; ix) it is also
assumed that the amount of the generated signals comply with the
EEG signals belonging to the system under study.

\section{MONTE CARLO SIMULATION AND RESULTS}

\begin{figure}[c] \epsfig{file=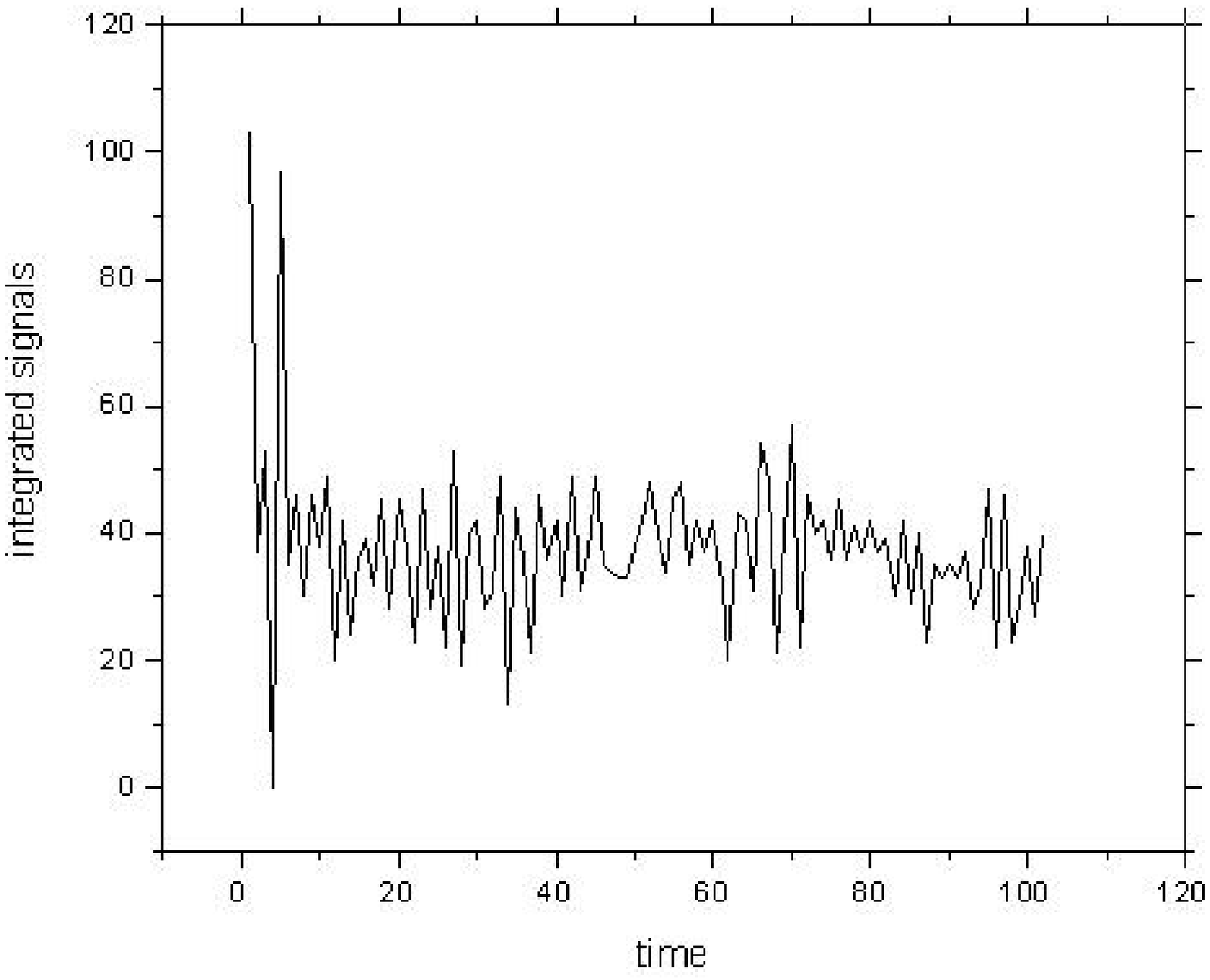, width=15cm, height=15cm}
\caption{Typical episodes of electrical activity between NL
elements similar to electrical activity of human brain as recorded
from EEG}\label{fig1}
\end{figure}

We have used 1000 NL elements to obtain integrated signals for
Monte Carlo (MC) simulation. The simulation is based on our model
mentioned above. The results exhibited EEG-like signals (see Fig.
\ref{fig1}). As seen in Fig. \ref{fig1}, these signals are not
periodic. Lyapunov exponents $\lambda$ for these signals were
found to be greater than zero, indicating that signals from NL
elements exhibit a chaotic behavior.

\begin{figure}[c] \epsfig{file=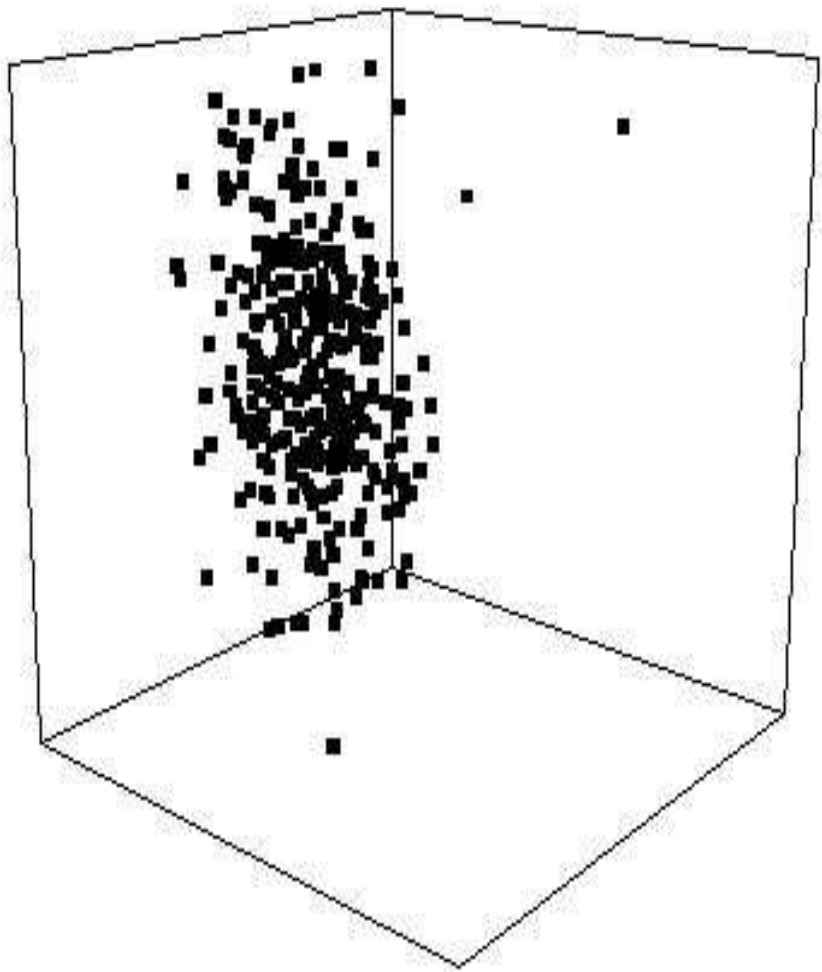, width=15cm, height=15cm}
\caption{Attractor of integrated signals at the phase space. This
portrait is the two- dimensional projection of the
three-dimensional construction : Demonstration of the plotted
integrated signals}\label{fig2a}
\end{figure}

\begin{figure}[c] \epsfig{file=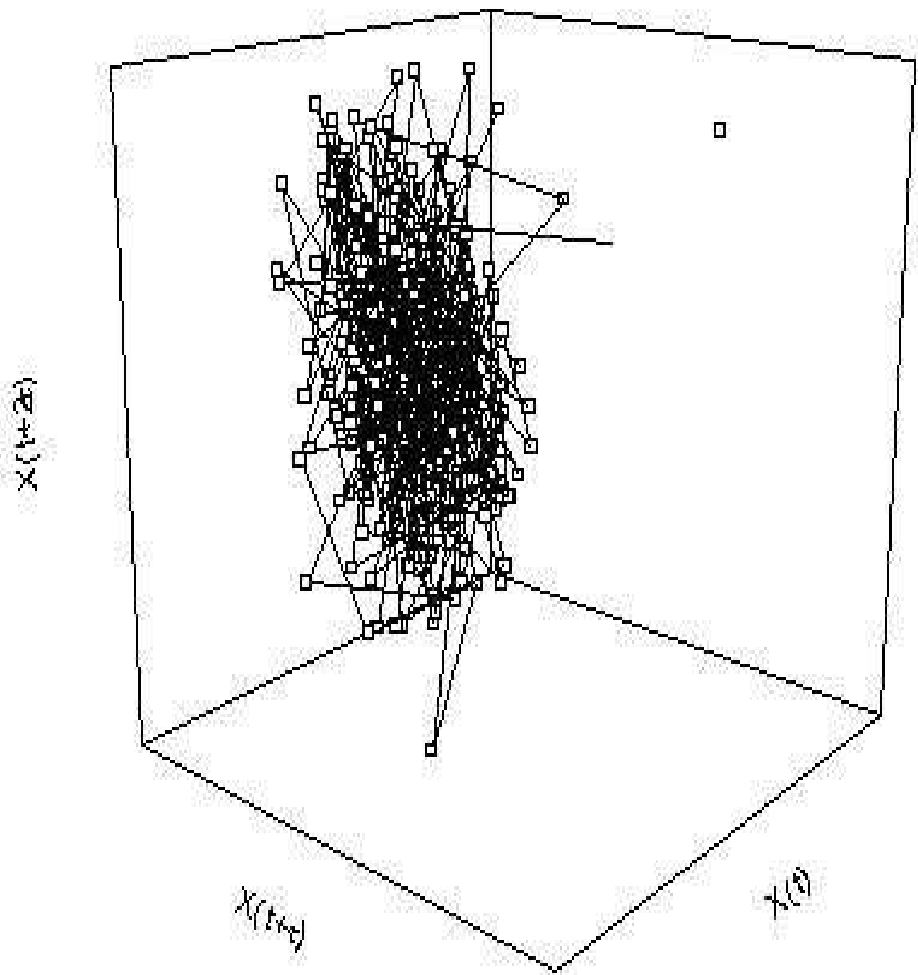, width=15cm, height=15cm}
\caption{Attractor of integrated signals at the phase space. This
portrait is the two- dimensional projection of the
three-dimensional construction : Demonstration of the trajectory
of these integrated signals}\label{fig2b}
\end{figure}

Fig. \ref{fig2a} and Fig. \ref{fig2b} illustrates the phase
portraits of the integrated signals in the phase space. To
construct phase space, we have used three dimensions, chosen as
$x(t)$, $x(t + \tau )$, and $x(t + 2\tau )$, with $\tau=1$. These
phase portraits are similar to the EEG phase portraits of
experimental data from others \cite{8,16}.

\begin{figure}[c] \epsfig{file=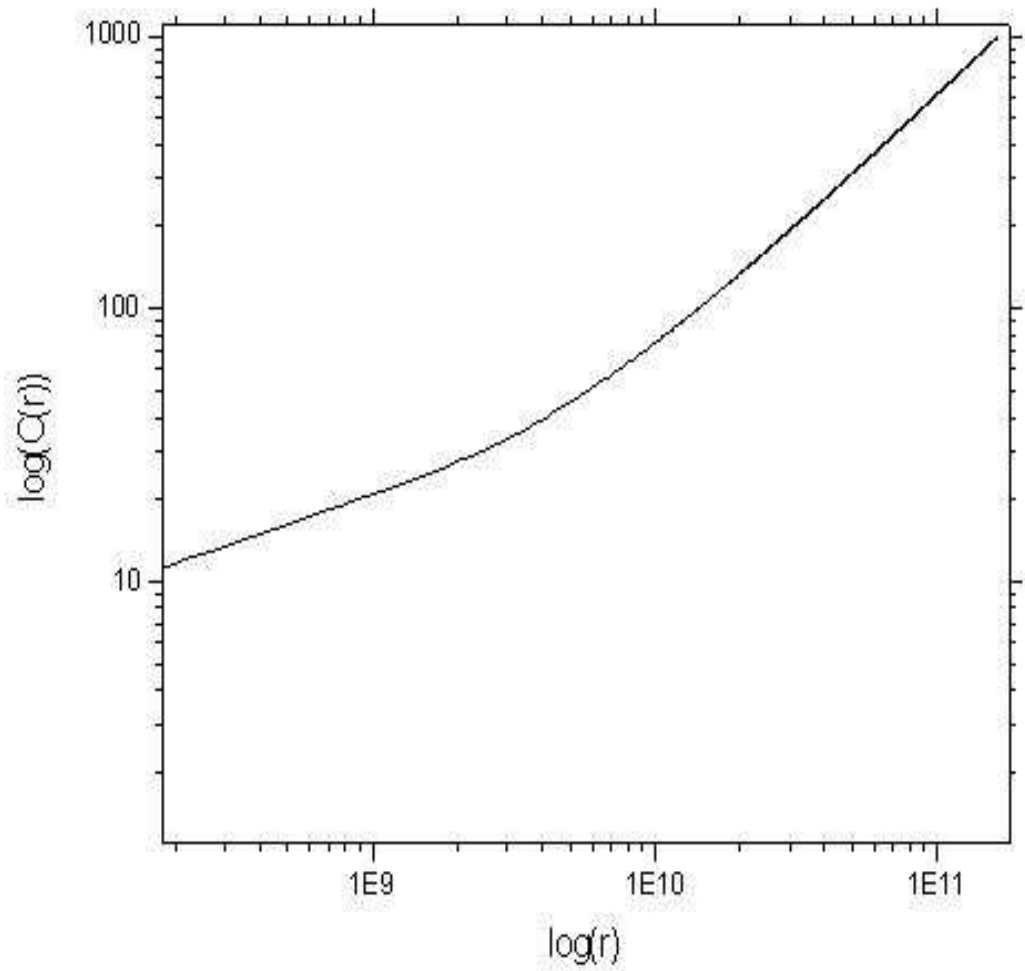, width=15cm, height=15cm}
\caption{Log-Log plot of $C(r)$ versus $r$ gives the correlation
dimension $D_2$}\label{fig3}
\end{figure}

\begin{figure}[c] \epsfig{file=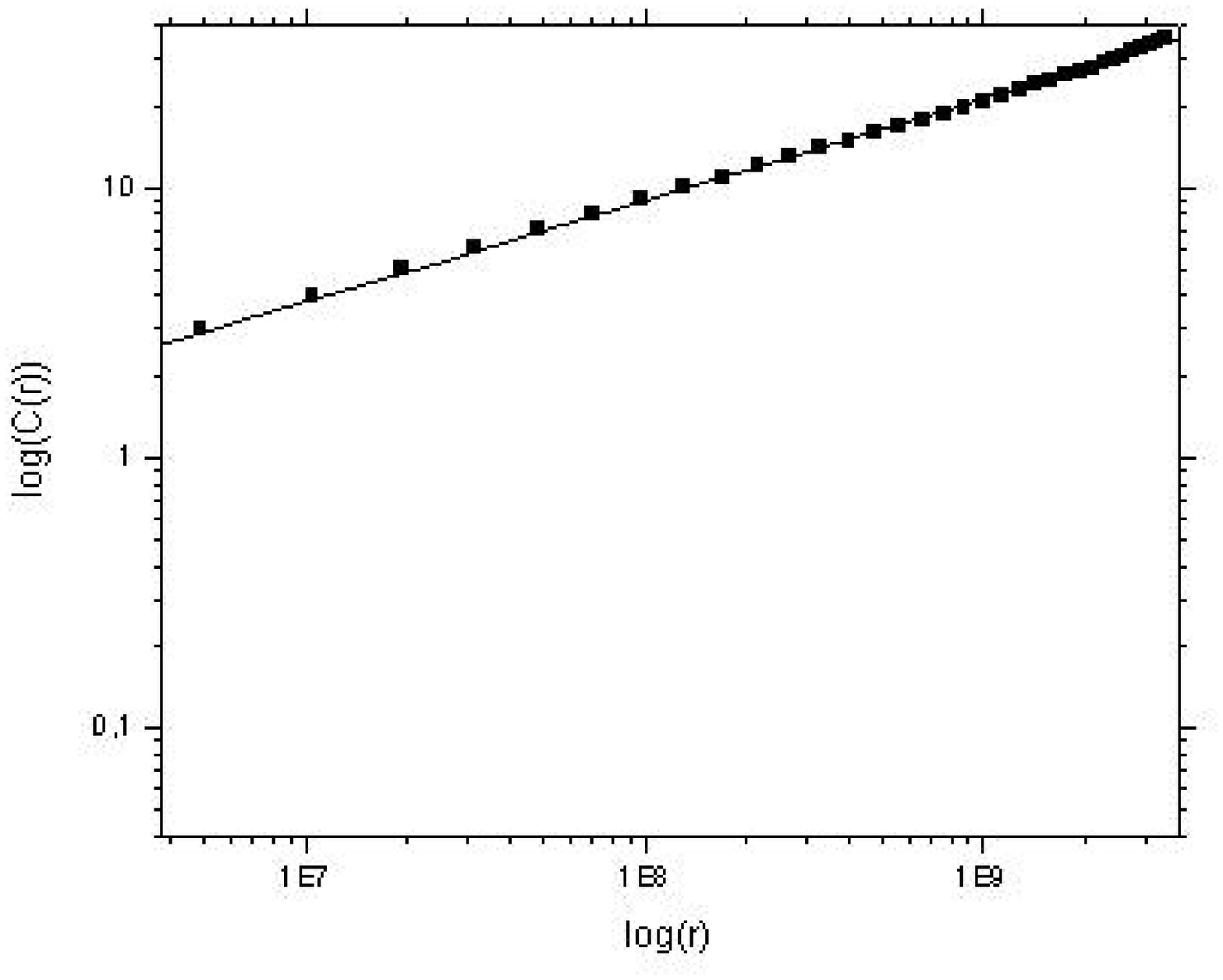, width=15cm, height=15cm}
\caption{Slopes of $\log (C(r))$  versus $\log (r)$ give the
correlation dimensions corresponding to the first
regime}\label{fig4a}
\end{figure}

\begin{figure}[c] \epsfig{file=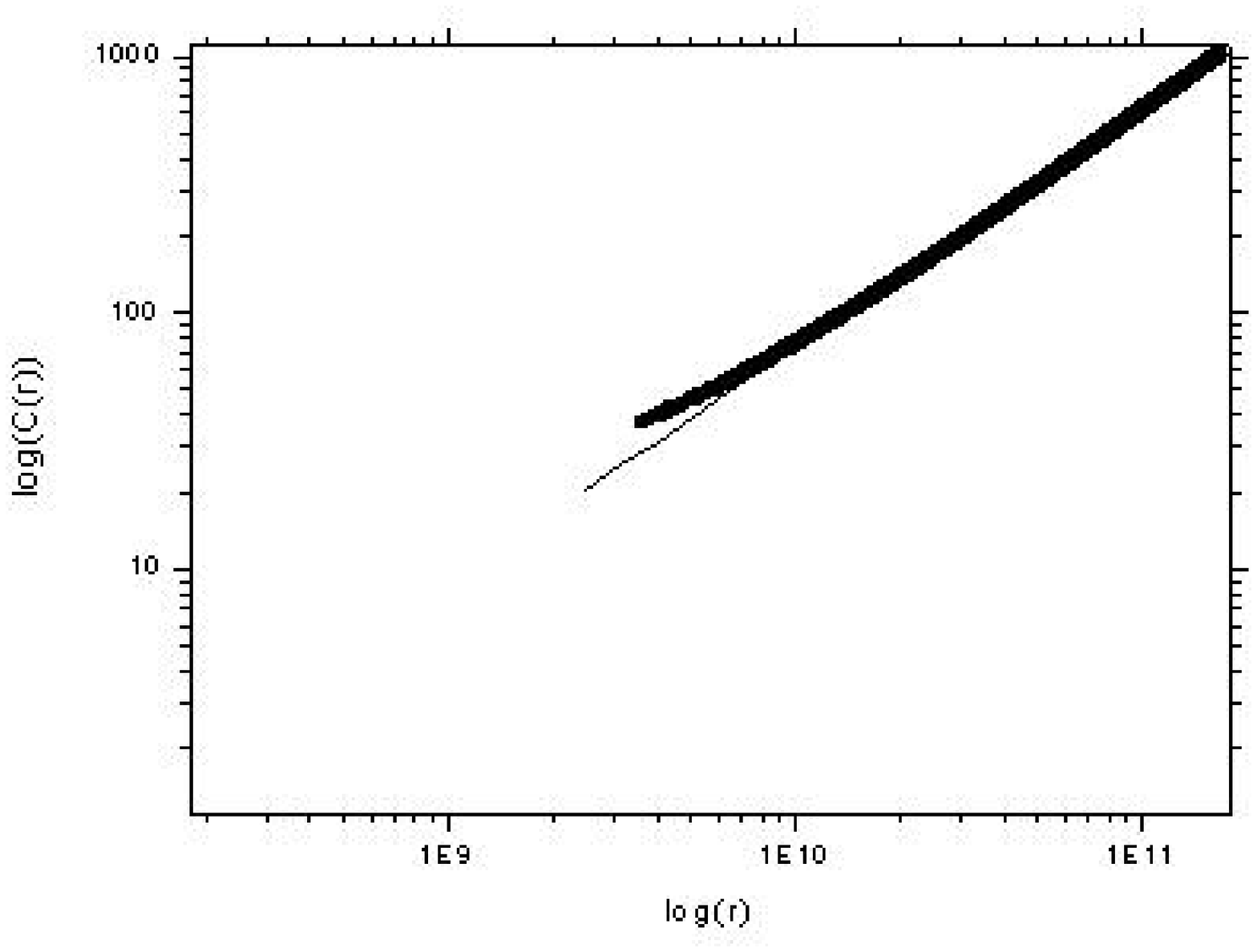, width=15cm, height=15cm}
\caption{Slopes of $\log (C(r))$  versus $\log (r)$ give the
correlation dimensions corresponding to the second
regime}\label{fig4b}
\end{figure}

A correlation dimension, known as fractal dimension in the
scientific literature, is a measure of chaotic behavior of the
attractor. The correlation dimension $D_2$ is given by (\ref{4})
Therefore, the slope of the curve ($\log C(r)$ against $\log r$)
gives $D_2$. The curve exhibits two different regimes as seen in
Figures \ref{fig3} and \ref{fig4a}, \ref{fig4b}. In the first
regime, the slope of the curve is 0.37, in the second regime, the
slope of the curve is 0.92, i.e., corresponding $D_2$ values are
0.37 and 0.92, respectively.

\section{CONCLUSION AND DISCUSSION}

In the present work, we have constructed a simple neuron model. If
the model was  solved using the well known MC simulation
technique, the results yielded EEG-like signals. The Lyapunov
exponent was found to be positive, and hence these EEG-like
signals had chaotic behaviors. The obtained EEG signals were put
on a three dimensional phase space, which gave us attractors.
However, these chaotic attractors constructed in phase space could
not be clearly seen, contrary to the attractors obtained from the
solution of the differential equations such as Lorentz type. The
reason may be a large number of attractors screen each other. The
attractors were similar to EEG normally recorded from the human
scalp. The very large number of attractors suggests that the
system would have many stable states changing with time. We have
also evaluated the correlation dimension, $D_2$ , corresponding to
the attractors (see above).  Our data indicate two regimes, for
them $D_2$ values were found to be 0.37 and 0.92, respectively.
The regime with $D_2=0.37$ corresponds to a very narrow time
interval at small times; the regime with $D_2=0.92$ seems to
correspond to a rather large time interval. Thus one can assume
that for the derived EEG-like signals $D_2 \cong 0.92$. However,
this value is inconsistent with others (see \cite{30}).
Accordingly, $D_2$ values were  reported to be $4.05 \pm 0.50$ and
$2.05 \pm 0.09$ in deep sleep and epilepsy, respectively \cite{5}.
Apparently, the model is open for further evaluations to
understand the stochastic brain dynamics. Different
psycho-physiological states in different subjects may be the
reason for the contradictory results mentioned above. We have
concluded that our tentative model constructed to simulate the
chaotic brain activity as seen in EEG may be useful in analyzing
the system dynamics of thousands of  cerebral neurons interacting
with each other.

\section*{Figure captions}
\begin{itemize}

\item[]{\bf Figure 1}:Typical episodes of electrical activity
between NL elements similar to electrical activity of human brain
as recorded from EEG.

\item[]{\bf Figure 2}:Attractor of integrated signals at the phase
space. This portrait is the two- dimensional projection of the
three-dimensional construction : Demonstration of the plotted
integrated signals

\item[]{\bf Figure 3}:Attractor of integrated signals at the phase
space. This portrait is the two- dimensional projection of the
three-dimensional construction : Demonstration of the trajectory
of these integrated signals.

\item[]{\bf Figure 4}:Log-Log plot of $C(r)$ versus $r$ gives the
correlation dimension $D_2$.

\item[]{\bf Figure 5}:Slopes of $\log (C(r))$  versus $\log (r)$
give the correlation dimensions corresponding to the first regime.

\item[]{\bf Figure 6}:Slopes of $\log (C(r))$  versus $\log (r)$
give the correlation dimensions corresponding to the second regime
.

\end{itemize}

\section*{References}
\begin{thebibliography}{99}
\bibitem[1]{1}
Achermann, P., Hartmann, R., Gunzinger, A., Guggenbühl, W. \&
Borberly, A.A., \emph{Correlation dimension of the human sleep
electroencephalogram: cyclic changes in the course of the night.
},
\newblock The European journal of neuroscience {\bf 6}, 497-500
(1994).


\bibitem[2]{2}
Aihara, K., Takabe, T., \& Toyoda, M., \emph{Chaotic neural
networks},
\newblock Physics Letters A, {\bf 144}, 333-340 (1990).

\bibitem[3]{3}
Aihara, K.,
\newblock In: The handbook of the brain theory and neural networks,
(p.183), ed. Arbib M.A., MIT press, Cambridge.(1995).


\bibitem[4]{4}
Amit, D.J.,
\newblock Modelling brain function, (pp.13-14). Cambridge University
Press.(1989).


\bibitem[5]{5}
Babloyantz, A. \& Destexhe,A., \emph{Low dimensional chaos in an
instance of Epilepsy.}, \newblock Proceeding of the national
academy of sciences of the USA, {\bf 83}, 3515-3517 (1986).

\bibitem[6]{6}
Baker, G.L. \& Gollup, J.P.,
\newblock Chaotic dynamics, (p. 144). Cambridge University
Press, (1996).

\bibitem[7]{7}
Bracewell, R.N.
\newblock The Fourier transform and its applications, (pp.1-65),
Mc Graw Hill., (2000).

\bibitem[8]{8}
Carnevale, N.T. \& Rosenthal, S., \emph{Kinetics of diffusion in a
spherical cell. I. No solute buffering}, \newblock Journal of
neuroscience methods {\bf 41}, 205-216 (1992).


\bibitem[9]{9}
Fitzhugh, R., \emph{Impulses and physiological states in
theoretical models of nerve Membrane}, \newblock Biophysical
journal {\bf 1}, 445-466 (1961).

\bibitem[10]{10}
Freeman, W.J., \emph{Simulation of chaotic EEG patterns with a
dynamic model of the olfactory systems}, \newblock Biological
cybernetics {\bf 56}, 139-150 (1987).

\bibitem[11]{11}
Gerstner, W., Ritz, R. \& van Hemmen, J.L., \emph{A biological
motivated and analytically soluble model of collective
oscillations in the cortex. I. Theory of weak locking}, \newblock
Biological cybernetics {\bf 68}, 363-374 (1993).

\bibitem[12]{12}
Grassberger, P. \& Procaccia, I., \emph{Measuring the strangeness
of strange attractors}, \newblock Physica D: Nonlinear phenomena
{\bf 9}, 189-208 (1983).

\bibitem[13]{13}
Hansel, D. \& Sompolinsky, H., \emph{Chaos and synchrony in a
model of a hypercolumn in visual cortex}, \newblock Journal of
computational neuroscience {\bf 3}, 7-34 (1996).

\bibitem[14]{14}
Hertz, J.A., Krogh, A. \& Palmer, R.G.,
\newblock Introduction to the theory of neural computation.
Addison-Wesley, Redwood City. (1991).


\bibitem[15]{15}
Hilborn, R.C.,
\newblock Chaos and nonlinear dynamics, (pp.407-409). Oxford
University Press (1994).


\bibitem[16]{16}
Hines, M.L. \& Carnevale, N.T., \emph{The NEURON simulation
environment},
\newblock Neural comptutation {\bf 9}, 1179-1209 (1997).

\bibitem[17]{17}
Hodgkin, A.L. \& Huxley, A.F., \emph{A quantitative description of
membrane current and its application to conduction and excitation
in nerve}, \newblock Journal of physiology {\bf 117}, 500-544
(1952).

\bibitem[18]{18}
Horn, D., \emph{Networks of complex neurons}, \newblock Physica A:
Statistical and theoretical physics {\bf 200}, 594-601 (1993).


\bibitem[19]{19}
Kaneko, K., \emph{Relevance of dynamic clustering to biological
networks}, \newblock Physica D: Nonlinear phenomena {\bf 75},
55-73 (1994).

\bibitem[20]{20}
Kattler, H., Dijk, D.-J. \& Borberly, A.A., \emph{Effect of
unilateral somatosensory simulation prior to sleep on the sleep
EEG in humans}, \newblock Journal of sleep research {\bf 3},
159-164 (1994).

\bibitem[21]{21}
Kinouchi, O. \& Tragtenberg, M.H.R., \emph{Modeling neurons by
simple maps}, \newblock International Journal of Bifurcation and
Chaos {\bf 6}, 2343-2360 (1996).


\bibitem[22]{22}
Kramarenko AV. \& Tan, U., \emph{Validity of spectral analysis of
evoked potentials in brain research}, \newblock International
journal of neuroscience {\bf 112}, 489-499 (2002).


\bibitem[23]{23}
Lai, Y.C., Osorio, I., Harrison, M.A.F. \& Frei, M.G.,
\emph{Correlation-dimension and autocorrelation fluctuations in
epileptic seizure dynamics},
\newblock Physical Review E {\bf 65}, art.no. 031921 (2002).

\bibitem[24]{24}
McCulloch, W.S. \& Pitts, W.A., \emph{A logical calculus of the
ideas immanent in neural nets}, \newblock Bulltein of mathematical
biophysics {\bf 5}, 115 (1943).

\bibitem[25]{25}
Nagumo, J., Arimoto, S. \& Yoshizawa, S., \emph{An active pulse
transmission line simulating 1214-nerve axons}, \newblock
Proceedings of Institute of Radio Engineers {\bf 50}, 2061-2070
(1960).

\bibitem[26]{26}
Nayfeh , A.H., \& Balachandran, B.,
\newblock Applied nonlinear dynamics. John Wiley \& Sons. (1994).


\bibitem[27]{27}
Osorio, I., Harrison, M.A.F., Lai, Y.C. \& Frei, M.G.,
\emph{Observation on the application of the correlation dimension
and correlation integral to the prediction of seizures},
\newblock Journal of clinical neurophysiology {\bf 18(3)}, 269-274
(2001).

\bibitem[28]{28}
Pasemann, F., \emph{A simple chaotic neuron}, \newblock Physica D:
Nonlinear phenomena {\bf 104}, 205-211 (1997).


\bibitem[29]{29}
Takens, F.,
\newblock In: Lecture notes in mathematics, eds. Rand D.A. \&
Young L.S., Springer- Verlag, Berlin. (1981).

\bibitem[30]{30}
West, B.J., Novaes, M.N. \& Kavcic, V.,
\newblock In: Fractal geometry in biological systems (pp.
267-317), eds. Iannaccone P.M. \& Khokha M., CRC Press. (1996).


\end{thebibliography}
\end{document}